\documentclass[aip,graphicx, reprint]{revtex4-1}
\draft 
\usepackage{graphicx}
\usepackage{dcolumn}
\usepackage{bm}
\usepackage[english]{babel}
\usepackage[utf8]{inputenc}
\usepackage[T1]{fontenc}
\usepackage{mathptmx}
\usepackage{amsmath}
\usepackage{amssymb}

\begin{document}


\title{Calculating RF current condensation with self-consistent ray-tracing} 



\author{R. Nies}
\email[]{rnies@pppl.gov}

\author{A. H. Reiman}
\email[]{reiman@pppl.gov}

\author{E. Rodriguez}
\email[]{eduardor@princeton.edu}

\affiliation{Department of Astrophysical Sciences, Princeton University, Princeton, NJ, 08543}
\affiliation{Princeton Plasma Physics Laboratory, Princeton, NJ, 08540}

\author{N. Bertelli}
\email[]{nbertell@pppl.gov}
\affiliation{Princeton Plasma Physics Laboratory, Princeton, NJ, 08540}

\author{N. J. Fisch}
\email[]{fisch@princeton.edu}

\affiliation{Department of Astrophysical Sciences, Princeton University, Princeton, NJ, 08543}
\affiliation{Princeton Plasma Physics Laboratory, Princeton, NJ, 08540}

\date{\today}

\begin{abstract}
By exploiting the nonlinear amplification of the power deposition of RF waves, current condensation promises new pathways to the stabilisation of magnetic islands. We present a numerical analysis of current condensation, coupling a geometrical optics treatment of wave propagation and damping to a thermal diffusion equation solver in the island. Taking into account the island geometry and relativistic damping, previous analytical theory can be made more precise and specific scenarios can be realistically predicted. With this more precise description, bifurcations and associated hysteresis effects could be obtained in an ITER-like scenario at realistic parameter values. Moreover, it is shown that dynamically varying the RF wave launching angles can lead to hysteresis and help to avoid the nonlinear shadowing effect.
\end{abstract}


\maketitle 


\section{Introduction}
Reliable mitigation and avoidance of disruptions is critical to the success of ITER and potential future tokamak power plants. Sudden loss of plasma confinement poses a serious threat to machine components through high heat loads, EM forces and runaway electrons \cite{boozer_theory_2012, boozer_magnetic_2019, lehnen_disruptions_2015}. In the JET tokamak equipped with an ITER-like wall, 95\% of natural disruptions are preceded by magnetic islands \cite{gerasimov_overview_2018}, making their stabilisation an essential task.

Magnetic islands are suppressed by generating a stabilising resonant component of the magnetic field, as is typically done by driving current at the island O-point. Current is generally driven directly by RF waves \cite{fisch_confining_1978,fisch_creating_1980,fisch_current_1981, karney_currents_1981, karney_efficiency_1985,karney_comparison_1985, bonoli_simulation_1986,fisch_theory_1987,karney_greens_1989,decker_eccd_2003,lin-liu_electron_2003,prater_heating_2004}, such as electron-cyclotron (EC) and lower-hybrid (LH) waves. Additionally, by depositing power and thereby heating the plasma, RF waves can also modify the ohmic current profile by decreasing the local resistivity. Thus, both RF heating and current drive can be used to stabilise magnetic islands, as has been investigated theoretically \cite{chan_stabilization_1982, reiman_suppression_1983, hegna_stabilization_1997, yu_modeling_1999, de_lazzari_merits_2009, bertelli_requirements_2011, pratt_early_2016} and experimentally \cite{hoshino_avoidance_1992, kislov_m2_1997, gantenbein_complete_2000, isayama_complete_2000, zohm_physics_2001, zohm_neoclassical_2001, haye_control_2002, the_textor_team_effect_2007,westerhof_tearing_2007,zohm_control_2007,maraschek_control_2012, kolemen_state---art_2014}. For instance, stabilisation of magnetic islands with ECCD is planned in ITER \cite{ramponi_iter_2007, figini_assessment_2015, poli_electron_2018}.

Typically, both EC and LH waves are damped on fast superthermal electrons. Therefore, the damping rate strongly depends on temperature through the electron population in the tail of the distribution function. This high sensitivity of power deposition to temperature can result in a positive feedback loop, where the magnetic island is heated by the RF wave, the elevated temperature leads to an increased power deposition, and so forth. This nonlinear effect, called current condensation\cite{reiman_suppression_2018, rodriguez_rf_2019}, can lend further help in stabilising islands, as narrower power deposition and current profiles centred on the island's O-point can be achieved\cite{reiman_suppression_2018, rodriguez_rf_2019}. Furthermore, current condensation can lead to bifurcations, where the island temperature would increase without bound, if not for other limiting effects such as depletion of the wave power \cite{rodriguez_rf_2019} or stiff temperature gradients \cite{rodriguez_rf_2020}. 

The present work extends previous analytical studies \cite{reiman_suppression_2018, rodriguez_rf_2019, rodriguez_rf_2020} with a numerical approach of current condensation. A geometrical optics treatment of wave propagation and damping is coupled with a solver of the thermal diffusion equation in the island geometry, as presented in Sec.~\ref{sec:occami}. The calculation iterates between solution of the thermal diffusion equation and calculation of the power deposition along the ray trajectories in the presence of the perturbed temperature to obtain a self-consistent solution of the nonlinear coupled system.

The geometrical optics approach allows for the inclusion of relativistic effects in the damping. We show in Sec.~\ref{sec:relativistic_damping} that the previously developed theory of current condensation \cite{reiman_suppression_2018, rodriguez_rf_2019} can be generalised to account for these relativistic effects, as well as the island geometry. In Sec.~\ref{sec:ITER_bifurcation}, we obtain values and trends in the bifurcation threshold consistent with previous work. Furthermore, the same calculations show that a bifurcation and associated hysteresis in the island temperature could be obtained in ITER-like H-mode and L-mode scenarios, at realistic values of input power, diffusion coefficient and island temperature perturbation. A constant diffusion coefficient was however used, an approximation which holds for low temperature perturbations, below the ITG instability threshold. An estimate suggests that this is indeed justified for the lowest temperature perturbations we observed at a bifurcation, although more detailed calculations will need to be performed in the future.

Finally, we show in Sec.~\ref{sec:hysteresis_launching_angle} that a bifurcation and hysteresis can be obtained by varying the RF wave launching angles, which could be a pathway for future experimental verification of current condensation. Additionally, the launching angles can be adjusted to circumvent the nonlinear shadowing effect \cite{rodriguez_rf_2019, jin_pulsed_2020}.

\section{Coupling of ray-tracing and magnetic island model}
\label{sec:occami}

The numerical approach presented below aims to simulate the nonlinear dynamics of current condensation, yielding self-consistent temperature and power depositions. The newly developed code OCCAMI (Of Current Condensation Amid Magnetic Islands) couples the ray-tracing code GENRAY \cite{smirnov_calculations_1995} with a driven heat diffusion equation solver for the magnetic island. The ray-tracing computes the RF wave propagation and damping. The ensuing power deposition is used to solve the steady-state diffusion equation to obtain the temperature profile in the island. The temperature is then given back as an input to the ray-tracing code. This process is repeated until convergence in the island temperature is attained. 

The temperature and power deposition thus obtained are self-consistent and allow us to investigate current condensation. This numerical treatment expands on previous analytical work where the initial power deposition was assumed to be constant \cite{reiman_suppression_2018} or exponentially decreasing around a peak \cite{rodriguez_rf_2019}, although the latter also incorporated self-consistent depletion of the wave energy. Furthermore, geometric effects inherent to magnetic islands are now included in the heat diffusion equation solver, whereas a slab model had previously been used to keep the problem analytically tractable \cite{rodriguez_rf_2019}. 

We now present in more detail the coupling of the ray-tracing for wave propagation and absorption with the heat diffusion equation solver.

\subsection{Ray-tracing for wave propagation and power deposition}

The code GENRAY \cite{smirnov_calculations_1995} simulates the propagation and absorbtion of electromagnetic waves in the geometrical optics approximation. The coupling of the island model with the ray-tracing calculations occurs through the temperature profile. We are therefore assuming that the perturbation in the magnetic field $B_r$ associated with a magnetic island is small, with a negligible impact on ray propagation and absorption.

\begin{figure}
\includegraphics[width=0.5\textwidth]{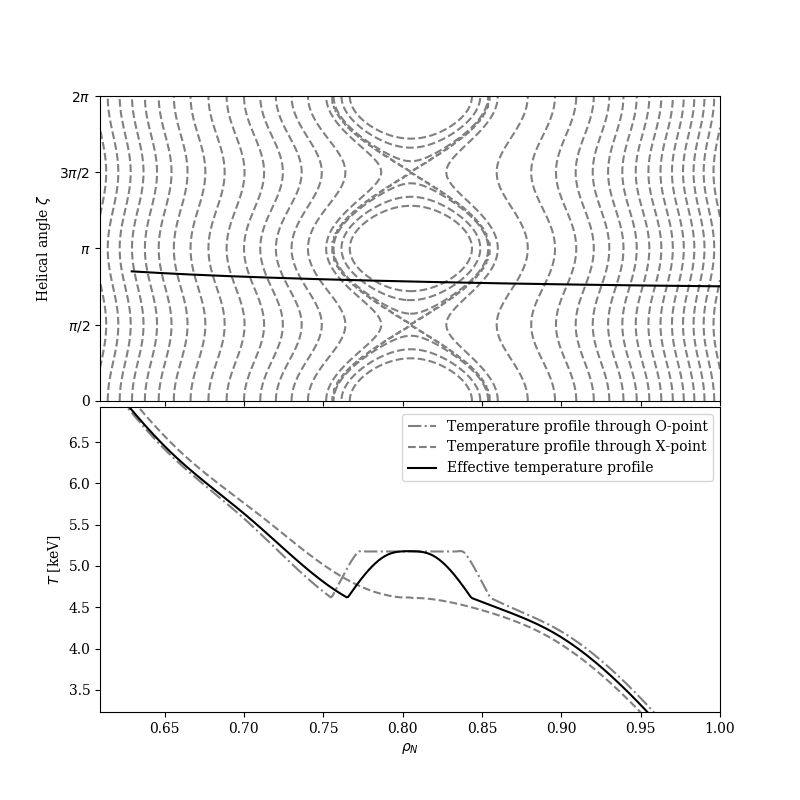}
\caption{Effective Temperature profile input to GENRAY, as a function of normalised radius $\rho_N = \sqrt{\psi_t}$, with the toroidal flux $\psi_t$. The upper plot shows the 2-dimensional temperature profile with a $(M=2,N=1)$ magnetic island. The dotted lines are temperature contours, while the solid line represents the ray trajectory. The lower plot shows different cuts in the upper plot: at $\zeta = 0,\pi$ (O-point), $\zeta = \pi/2, 3\pi/2$ (X-point) and the effective temperature profile for the shown ray trajectory. Note that the flat temperature profile through the O-point in the lower plot arises from the fact that there is no power deposited in the central flux surfaces in the island in this case.}
\label{fig:island_effective_T}
\end{figure}

Axisymmetry of the plasma is lost in the presence of magnetic islands, whence the island temperature profile becomes a function of not only a radial coordinate, but also the helical angle $\zeta = \theta - N/M \varphi$, with the poloidal (toroidal) angles $\theta$ ($\varphi$) and poloidal (toroidal) mode numbers $M$ ($N$). However, the code GENRAY assumes axisymmetry, by requiring a one-dimensional temperature profile as input. We incorporate the island-linked asymmetry through an effective temperature profile for each ray. This effective profile corresponds to the temperature profile that the ray experiences as it propagates through the plasma, as illustrated in Fig.~\ref{fig:island_effective_T}. This assumes that the ray trajectory is not significantly altered by the change in temperature between iterations. In the case of multiple passes, i.e. when the ray trajectory traverses a given radius at multiple points in its trajectory, the effective profile is taken to be that experienced by the ray in its last traversal of the island. This is a reasonable approximation when most of the power is absorbed in a single pass, i.e. we assume the damping on all passes but the last must have been negligible. We will consider EC waves in this study, for which this is generally the case.

The solution for the temperature shown in Fig.~\ref{fig:island_effective_T} corresponds to a case where there is a single ray propagating through a locked island.  For a rotating island, it is necessary to calculate the total power deposited in the island through one rotation. This has been implemented for a fast rotating island, where the diffusion time is much longer than the island rotation time. Then, the total power deposition can be approximated by averaging the power deposition along multiple ray trajectories sampling the island at different phases.

\subsection{Heat diffusion equation solver in island geometry}
\label{sec:diff_eq}

The power deposition obtained from the ray-tracing calculation is used to update the island temperature. Integrating the steady-state diffusion equation once (a detailed derivation can be found in appendix~\ref{sec:derivation_diff_eq}), we obtain
\begin{equation}
    \frac{\partial u}{\partial \sigma} = - \frac{P_\text{dep}(\sigma)}{n\chi_\perp T_s} 
    \frac{\sigma}{E(\sigma) - \left(1-\sigma^2\right) K(\sigma)} \frac{W M}{32\pi\;r_r R_0}. \label{eq:diffusion}
\end{equation}
Here, $u = \left( T-T_s\right)/T_s$ is the normalised island temperature, with the temperature at the separatrix $T_s$, $P_\text{dep}(\sigma)$ is the power deposited within the island flux surface $\sigma$ ($\sigma = 0, 1$ at the O-point and separatrix, respectively), $E(\sigma)$ and $K(\sigma)$ are respectively the complete elliptical integrals of the first and second kind, $W$ is the island width, $r_r$ is the minor radius at the resonant surface, $R_0$ the tokamak major radius and $M$ the island's poloidal mode number. The temperature at the separatrix $T_s$ is assumed to be constant in our simulations. This corresponds to the case where, for example, the EC power is initially deposited radially inward from the island, at radii $r<r_r-W/2$, and then redirected outward to $r \sim r_r$. As the total power deposited within the flux surface at $r = r_r + W/2$ remains identical, the temperature gradient at radii $r \geq r_r + W/2$ is unchanged, and so the separatrix temperature is constant \cite{rodriguez_rf_2019}. More generally, the temperature perturbation at the separatrix may be negligibly small, but this is not always the case.

Furthermore, the perpendicular heat diffusion coefficient $\chi_\perp$ is taken to be constant in this study. The interplay between current condensation effects and a variable heat diffusion coefficient in the form of stiff gradients was investigated analytically in \citet{rodriguez_rf_2020}. A corresponding numerical treatment with OCCAMI is left for future work.

As it stands, the diffusion equation has been reduced to a 1st order ODE (Eq.~\ref{eq:diffusion}). It can thus be readily solved for the island temperature profile, given the power deposited from GENRAY and the boundary condition ${u(\sigma=1) = 0}$. This is done numerically with a 4th order Runge-Kutta integrator. Note that the second boundary condition for the original 2nd order diffusion equation, ${\partial u / \partial \sigma (\sigma = 0) = 0}$, was used when integrating the originally 2nd order diffusion equation, and is required for regularity.

The obtained temperature is then fed back to GENRAY through an updated effective temperature profile and the ray propagation and power deposition are calculated anew. This cycle is repeated until convergence in the island temperature is reached, i.e. when the relative change in the normalised temperature perturbation between iterations is below a given threshold $\epsilon$ (in this study, $\epsilon = 5\cdot 10^{-4}$ was chosen).

\section{Effects of relativistic damping}
\label{sec:relativistic_damping}

In this section, we show how the sensitivity of damping to temperature in the classical case, $w_0$, can be generalised to account for relativistic damping effects, leading to the definition of an effective $w_\text{eff}$. Approximate formulas for the O1 and X2 modes are presented. Spatial variation of the damping within the island must also be taken into account, leading to the introduction of an average $\overline{w}_\text{eff}$. To keep this study self-consistent and motivate our analysis of $w_\text{eff}$, we first present a brief summary of the linear theory of resonant wave damping, and of the basic theory of relativistic effects in electron cyclotron wave damping.

\subsection{Linear theory of resonant wave damping}
\label{sec:linear_theory_validity}

In addition to calculating the wave propagation, ray-tracing codes such as  GENRAY\cite{smirnov_calculations_1995} also generally provide the damping coefficient of the wave, obtained from the anti-hermitian part of the dielectric tensor. Formulas for the dielectric tensor in various regimes are obtained in the linear regime of wave damping, i.e. assuming the distribution function to be Maxwellian. However, this is only an approximation as the distribution function is modified by the wave interaction through quasi-linear diffusion of particles in velocity space, making the absorption of RF wave energy a nonlinear process. Nevertheless, the linear theory is a suitable approximation in multiple scenarios, as shown below, and we thus make use of it in the analysis and simulations presented in this study.

In certain cases, e.g. at low wave power, the electron distribution remaining Maxwellian is a valid assumption. Moreover, for EC waves, the linear theory was found to be valid and nearly independent of RF wave power, in the non-relativistic limit and for diffusion in the perpendicular velocity $v_\perp$ only. The classical (non-relativistic) resonance condition between EC wave and electrons is $k_\parallel v_\parallel = \omega - n \Omega$, with $n$ the harmonic number of the resonance, $k_{\parallel}$ the wavenumber parallel to the magnetic field, $\omega$ the wave frequency, $\Omega = eB/m_e$ the cyclotron frequency and $v_\parallel$ the parallel velocity of resonant electrons. Then, if the particles diffuse in $v_\perp$ only, they remain in resonance with the wave as the classical resonance condition is independent of $v_\perp$\cite{karney_currents_1981}. The analysis will be more complicated in the relativistic case, where the resonance condition depends on $v_\perp$ (see below, Eq.~\ref{eq:relativistic_resonance}), or when diffusion of particles also occurs in the $v_\parallel$ direction. 

Evidence for the linear theory's more general validity can be found in a study comparing the current drive obtained from linear theory with results of Fokker-Planck calculations, which solve the nonlinear damping problem, and with experiment: good agreement was shown for a large range of parameters in DIII-D \cite{petty_detailed_2002}. The linear theory was also shown to yield accurate results for an ITER benchmark scenario in a study comparing multiple ray-tracing codes (including GENRAY), which compute the damping from linear theory, with two Fokker-Planck solvers \cite{prater_benchmarking_2008}. Nevertheless, additional Fokker Planck calculations need to be performed for a broader range of parameters spanning those investigated in our study.

\subsection{Relativistic resonance and damping}

As shown in Sec.~\ref{sec:linear_theory_validity}, the classical resonance condition between EC wave and electrons is independent of $v_\perp$. The spatial damping rate is obtained by integrating over the distribution function of resonant electrons in velocity space, and will thus be proportional to the population of electrons with parallel velocity satisfying the resonance condition. In the linear regime, i.e. when the distribution function in the parallel velocity is Maxwellian, the spatial damping rate of EC waves thus obeys $\alpha \propto e^{-w_0^2}$, with the thermal velocity $v_T$,  and $w_0 = (\omega - n\Omega)/(k_\parallel v_T)$ (e.g.~\citet{swanson_plasma_2012}).

However, relativistic effects on the damping cannot be neglected in realistic scenarios. Indeed, the classical resonance condition needs to be modified to take into account the relativistic mass increase (e.g.~\citet{fidone_role_1982}):
\begin{equation}
    \omega - n\frac{\Omega}{\gamma} = k_{\parallel} v_{\parallel}. \label{eq:relativistic_resonance}
\end{equation} 
In the relativistic case, the resonance follows an ellipse in $(v_{\parallel}, v_\perp)$ space, due to the factor $\gamma = (1 - ( v_{\parallel}^2 + v_\perp^2 )/c^2)^{-1/2}$, according to Eq.~\ref{eq:relativistic_resonance}. In particular, this sets an important constraint for resonance on the low-field side ($n\Omega < \omega$),
\begin{equation}
    \frac{n\Omega}{\omega} \geq \sqrt{1 - N_{\parallel}^2}, \label{eq:rel_constraint}
\end{equation}
where $N_{\parallel} = k_{\parallel} c / \omega$ is the parallel refractive index. As $\Omega \propto B \sim 1/R$, Eq.~\ref{eq:rel_constraint} leads to a relativistic boundary, rendering part of the tokamak's low-field side inaccessible to heating and current drive with EC waves. This is most apparent for low values of the parallel refractive index, for which the relativistic boundary is close to the resonance $\Omega/\omega=1$, and the wave typically damps very strongly in a narrow spatial region.

In the following, we present an approximate form of the damping coefficient $\alpha$ for the O1 mode, as is appropriate for ITER. The corresponding formulas for the X2 mode can be found in Appendix~\ref{sec:weff_X2_mode}. For the O1 mode, we assume absorption of the wave's L-polarisation to be negligible, and the wave's R-polarisation to be generally small. Then the damping is due mainly to the electric field component along the background magnetic field. In that case, we can approximate the damping coefficient as
\begin{equation}
    \alpha \approx | E_z / E | ^2 \frac{\omega}{c N} \epsilon_{33}^{\prime \prime}, \label{eq:damping_approx_eps}
\end{equation}
with the longitudinal polarisation $| E_z / E | $, the refractive index $N \approx \sqrt{1 - \omega_p^2/\omega^2}$, the electron plasma frequency $\omega_p = \sqrt{n_e e^2 / (\epsilon_0 m_e)}$ and electron density $n_e$. The anti-hermitian component $\epsilon_{33}^{\prime \prime}$ of the dielectric tensor, derived by \citet{fidone_role_1982} by integrating over the resonant ellipse in velocity space, is reproduced here:
\begin{align}
    \epsilon_{33}^{\prime \prime} \approx &\frac{\pi \omega_{p}^{2}}{2 \Omega^{2}} \frac{R^{7 / 2}}{N_\parallel^{3 / 2}} \frac{N_{\perp}^{2} \mu S}{\left(1-N_\parallel^{2}\right)^{5 / 2}}\Bigg[I_{3 / 2}(\xi)\left(1+\frac{N_\parallel^{2} \Omega^{2}}{R^{2} \omega^{2}}\right) \nonumber \\
    & -2 I_{5 / 2}(\xi)\left(\frac{2}{\xi}+\frac{N_\parallel \Omega}{R \omega}\right)\Bigg] e^{\mu\left(1-\frac{\Omega / \omega}{1-N_\parallel^{2}}\right)}, \label{eq:eps33_O1_mode}
\end{align}
where $I_\nu(\xi)$ are the $\nu$-th order modified Bessel functions of the first kind,
\begin{align}
    \mu &= \frac{m_e c^2}{T}, \\ 
    R   &= \sqrt{ \left( \frac{\Omega}{\omega}\right)^2-1+N_{\parallel}^2},\\
    \xi &= \frac{N_{\parallel}R\mu}{1-N_{\parallel}^2}, \qquad\qquad \text{and} \\
    S   &= H\left( \left( \frac{\Omega}{\omega}\right)^2-1+N_{\parallel}^2 \right)
\end{align}
is a Heaviside function enforcing the relativistic constraint of Eq.~\ref{eq:rel_constraint}.

\subsection{Sensitivity of relativistic damping to temperature}

As seen above, the linear damping rate of EC waves satisfies $\alpha \propto e^{-w_0^2}$ in the classical limit, with $w_0^2 \propto 1/T$. This strong sensitivity of damping to temperature is essential for the current condensation effect, with nonlinear effects becoming potentially relevant for $w_0^2 \Delta T / T \gtrsim 0.5$ \cite{reiman_suppression_2018}. The quantity $w_0^2$ therefore provides a direct indicator of the sensitivity of damping to temperature in the classical limit.

We are interested in obtaining the sensitivity of damping to temperature taking into account relativistic effects in the damping. We define an effective $w_\text{eff}$ as 
\begin{equation}
    w_{\text{eff}}^2 = T\partial_T (\ln\alpha), \label{eq:def_weff}
\end{equation}
such that $w_\text{eff}$ measures the sensitivity of damping to temperature, in analogy to the classical case. Indeed, Eq.~\ref{eq:def_weff} can be viewed as a first order correction term in a Taylor expansion of $\ln(\alpha)$. Therefore, for small temperature perturbations, $w_\text{eff}^2 \Delta T / T$ indicates the strength of nonlinear effects, as the damping is approximately amplified by a factor $\exp(w_\text{eff}^2 \Delta T / T)$. Finite temperature perturbations are treated in Sec.~\ref{sec:sub_rel_damping_current_condensation} in the regime $w_\text{eff}^2 \propto 1/T$.

Note that the damping satisfies $\alpha \propto e^{-w_\text{eff}^2}$ only in the case where $w_\text{eff}^2 \propto 1/T$, i.e. when $w_\text{eff}$ possesses the same temperature dependency as $w_0$, as can be shown by integrating Eq.~\ref{eq:def_weff}. However, $w_\text{eff}^2 \propto 1/T$ does not necessarily imply $w_\text{eff} = w_0$.

Using the definition of $w_\text{eff}$ in Eq.~\ref{eq:def_weff}, the approximate form of the damping in Eq.~\ref{eq:damping_approx_eps} and assuming the longitudinal polarisation $|E_z/E|$ to be independent of temperature (cold plasma approximation), we obtain $w_{\text{eff}}^2 \approx T \partial_T \left( \ln( \epsilon_{33}^{\prime \prime} ) \right) = -\xi \partial_\xi \left( \ln( \epsilon_{33}^{\prime \prime} ) \right)$. Combined with Eq.~\ref{eq:eps33_O1_mode}, this yields
\begin{equation}
    -w_{\text{eff}}^2 = 1 + \mu \left( 1 - \frac{\Omega/\omega}{1-N_{\parallel}^2} \right) + F(\xi, a),  \label{eq:w_eff2_formula}
\end{equation}
where
\begin{align}
    & a   = \frac{N_{\parallel}\Omega}{R\omega}, \label{eq:def_a}\\
    & F(\xi, a) = \nonumber \\ 
    & \xi \frac{ I_{3/2}(\xi) \left(-\frac{5}{2\xi} + \frac{3a^2}{2\xi} - 2a\right) + I_{5/2}(\xi) \left( 1 + a^2 + \frac{14}{\xi^2} + \frac{5a}{\xi} \right)}{ I_{3/2}(\xi) \left( 1+a^2\right) -2 I_{5/2}(\xi) \left( \frac{2}{\xi} + a\right)} \label{eq:def_F_xi}.
\end{align}

The $w_\text{eff}$ of Eq.~\ref{eq:w_eff2_formula} reduces to $w_0 = (\omega - \Omega)/(k_{\parallel}v_T)$ in the classical limit, consisting of $N_{\parallel}^2 \gg T/(m_e c^2)$ and $N_{\parallel}^2 \gg | 1 - \left( \Omega / \omega \right)^2|,$\cite{fidone_role_1982} as well as $\mu | 1-\left( \Omega / \omega \right) | \gg 1$, as the damping coefficient $\alpha$ also reduces to the classical formula in the same limit (see Appendix~\ref{sec:temp_dependence_weff}).

Our ray-tracing calculations employ the more general approximation of the dielectric tensor for a relativistic electron plasma from \citet{mazzucato_damping_1998}, which adds higher harmonic corrections to the treatment of \citet{fidone_role_1982}. However, it is found that Eq.~\ref{eq:w_eff2_formula} agrees well with a numerical evaluation of Eq.~\ref{eq:def_weff} in situations of interest for ITER, as shown in Appendix~\ref{sec:validity_weff}.

The temperature dependence of $w_\text{eff}$ defined in Eq.~\ref{eq:w_eff2_formula} is nontrivial, in contrast to the classical case where $w_0^2 \propto 1/T$.  It is however shown in Appendix~\ref{sec:temp_dependence_weff} that $w_\text{eff}^2 \propto 1/T$ is a suitable approximation in the limit $\xi \gg 1$, and $ N_\parallel^2 \sim | 1-(\Omega/\omega)^2| $. If $ N_\parallel^2 \gg | 1-(\Omega/\omega)^2| $ however, $w_\text{eff}^2 \propto 1/T$ can still hold when $\mu (1-(\Omega/\omega)) \gg 1$. The first set of conditions are typically satisfied for reasonably high $N_\parallel$ and not too high temperatures (see~Fig.~\ref{fig:test_weff_T}), as the condition $ N_\parallel^2 \gg | 1-(\Omega/\omega)^2| $ is very restrictive. Therefore, multiple insights from the theory developed in the case of classical damping remain valid in the relativistic case, e.g. the temperature perturbations necessary to obtain a bifurcation, as shown in Sec.~\ref{sec:ITER_bifurcation}.

\subsection{Relativistic damping and current condensation}
\label{sec:sub_rel_damping_current_condensation}

The $w_\text{eff}$ derived above is now connected to the current condensation effect. We define the nonlinear amplification parameter $\Theta_{NL}$ as the logarithmic change in the damping rate $\alpha$ due to a change in temperature $\Delta T$ from an unperturbed temperature $T_0$,
\begin{equation}
   \Theta_{NL} \equiv \ln{\left( \frac{\alpha(T=T_0+\Delta T)}{\alpha(T=T_0)}\right)}. \label{eq:theta_NL_def}
\end{equation}
For small temperature perturbations, using Eq.~\ref{eq:def_weff}, we obtain $\Theta_{NL} = w_\text{eff}^2 \Delta T / T$, a result valid for arbitrary forms of $w_\text{eff}$. The case of finite temperature perturbations can be treated by assuming $w_\text{eff}^2 \propto 1/T$ is valid (see Appendix~\ref{sec:temp_dependence_weff}). Then, Eq.~\ref{eq:def_weff} can be integrated to obtain $\alpha = \alpha_0 e^{-w_\text{eff}^2}$. Further, defining $u \equiv \Delta T / T_0$, Eq.~\ref{eq:theta_NL_def} reduces to
\begin{align}
    \Theta_{NL} &= -w_\text{eff}^2 \left(T = T_0(1+u) \right) + w_\text{eff}^2 \left(T = T_0\right) \nonumber \\
    &= w_\text{eff}^2\left(T=T_0 \right) \frac{u}{1+u}.\label{eq:theta_NL}
\end{align}
As expected, this quantity reduces to $\Theta_{NL} = w_\text{eff}^2 u$ for small temperature perturbations $u \ll 1$, as was assumed in previous work \cite{reiman_suppression_2018, rodriguez_rf_2019}.

The nonlinear amplification parameter $\Theta_{NL}$ proves useful to ascertain whether nonlinear effects like current condensation can become relevant in a given scenario. Indeed, $\Theta_{NL} \gtrsim 0.5$ is a necessary, but not sufficient, condition for nonlinear effects to become significant, while the limit $\Theta_{NL} \rightarrow 0$ corresponds to linear behaviour. Then, as $w_\text{eff} (T=T_0)$ can be obtained from the wave damping in the unperturbed temperature profile, an approximate lower bound on the temperature perturbation $u$ necessary to observe nonlinear effects can be obtained from inverting Eq.~\ref{eq:theta_NL} without performing the full nonlinear calculation. This motivates the use of the approximated form in Eq.~\ref{eq:theta_NL} instead of inserting the full damping coefficient into Eq.~\ref{eq:theta_NL_def}, which could not be readily solved for the temperature perturbation. Henceforth in this study, we will consider $w_\text{eff}$ to be evaluated at the initial unperturbed temperature $T_0$.

Although $w_\text{eff}$ and $\Theta_{NL}$ are useful local quantities, the damping rate can vary significantly within a given magnetic island, especially for large island widths. Indeed, while we generally assume the initial island temperature to be flat, the quantities $\Omega/\omega$, $N_\parallel$ and thus also $w_\text{eff}$ will in general be functions of position within the island. Thus, a suitable island average of $w_\text{eff}$ must be found. We define
\begin{equation}
    \overline{w}_\text{eff}^2 \equiv -\ln{ \left( \langle\exp{\left( -w_\text{eff}^2 \right)}\rangle_{\sigma\leq 1} \right) },\label{eq:w_eff_avg}
\end{equation}
where $\langle f \rangle_{\sigma\leq 1}$ is the mean value of $f$ within the magnetic island ($\sigma \leq 1$), evaluated along the ray. This averaging is motivated by the damping having the form $\alpha = \alpha_0 \exp{\left(-w_\text{eff}^2\right)}$ when $w_\text{eff}^2 \propto 1/T$ (see~Appendix~\ref{sec:temp_dependence_weff} for region of validity).


\section{Bifurcations and hysteresis in ITER}
\label{sec:ITER_bifurcation}

Current condensation can lead to bifurcations, where the nonlinear amplification of temperature leads to a runaway effect for the temperature and power deposited in the island. The temperature continues increasing \cite{reiman_suppression_2018}, until another limiting physical mechanism comes into play, leading to saturation. For example, the wave may have deposited all of its power\cite{rodriguez_rf_2019}, or the temperature increase might be limited by stiff temperature gradients \cite{rodriguez_rf_2020}.

Depending on the case, the limiting effects might lead the temperature to either smoothly transition to higher values, or experience a discontinuous jump at the bifurcation\cite{rodriguez_rf_2019}. In the latter case, hysteresis phenomena can be observed, as the jump from low to high temperature will not occur at the same parameter values as that from high to low temperature. Hysteresis curves can be traced out e.g. by varying the RF input power \cite{reiman_suppression_2018, rodriguez_rf_2019} (as considered in this section), the island width \cite{reiman_suppression_2018, rodriguez_rf_2019} or the RF wave launching angles, as shown in Sec.~\ref{sec:hysteresis_launching_angle}.

It is shown in this section that bifurcations can be obtained in ITER-like scenarios at realistic parameter values of temperature perturbation, diffusion coefficient and input power. Furthermore, values of the bifurcation threshold are shown to be consistent with previous work.

\subsection{Simulations setup}

\begin{figure}
\includegraphics[width=0.5\textwidth]{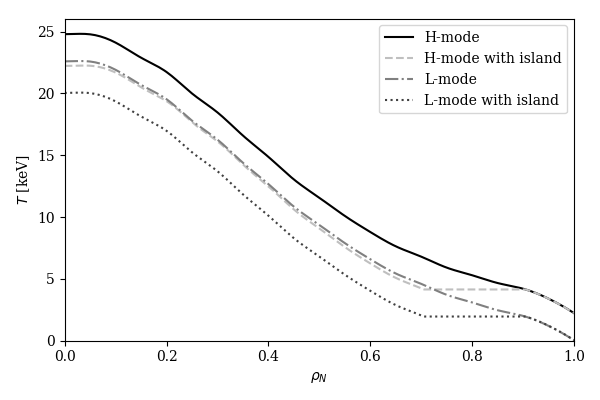}
\caption{Temperature profiles for H and L-mode scenarios, before and after local flattening due to the magnetic island ($W = 0.2, r_r = 0.805$). The flattened profiles are used as the initial unperturbed temperature profiles in our simulations. The L-mode temperature profile was obtained by substracting the pedestal height from the H-mode temperature profile. The temperature at the island separatrix is $T_s = 4.2$~and~$2.0$~keV for the H-mode and L-mode respectively. The H-mode profile closely resembles that of the ITER $15$~MA baseline scenario (e.g.~\citet{snicker_effect_2018}).}
\label{fig:temp_H_Lmode}
\end{figure}

The following simulations are based on ITER-like H- and L-mode scenarios, with temperature profiles shown in Fig.~\ref{fig:temp_H_Lmode}. A large island of width $W_N = 0.2$ (in units of the normalised radius $\rho_N = \sqrt{\psi_t}$, with the toroidal flux $\psi_t$) is introduced at the $q=2$ surface ($\rho_N = 0.805$), leading to a flattening of the temperature, as is also shown in Fig.~\ref{fig:temp_H_Lmode}. As the plasma is expected to fall back into L-mode for large island sizes due to deteriorated confinement, a pseudo L-mode scenario is also considered here, which was obtained by substracting the pedestal temperature from the H-mode profile (Fig.~\ref{fig:temp_H_Lmode}). A real L-mode profile would involve more substantial changes, e.g. to the magnetic equilibrium and density profiles. However, we will refer to our pseudo L-mode scenario simply as the L-mode scenario in the remainder of this study, a more detailed optimisation study for ITER being left for future work.

The diffusion coefficient in the island is assumed constant at $\chi_\perp = 0.1$~m$^2$~s$^{-1}$. Such small values are justified as turbulent transport is reduced due to the flattened temperature in the island region \cite{bardoczi_non-perturbative_2016, bardoczi_multi-field-scale_2017}. This approximation will thus break down when the Ion Temperature Gradient (ITG) threshold is exceeded, i.e. when
\begin{equation}
    \kappa_c \leq  -\frac{R}{T} \frac{\partial T}{\partial r} \approx \frac{R}{a}\frac{u_0}{W_N/2}, \label{eq:ITG_threshold}
\end{equation}
with the major radius $R = R_0 + r_r$, major radius at the magnetic axis $R_0=6.2$~m, minor radius $a=2.0$~m, normalised temperature perturbation at the island centre $u_0 = \Delta T(\sigma=0)/ T_s$, and ITG threshold $\kappa_c \approx 5$ for ITER \cite{garbet_profile_2004}. Then, effects of turbulent transport can be neglected when the temperature perturbations remain below $u_0/W_N \lesssim 1$, in which case the use of a low constant diffusion coefficient is justified. The maximally allowed temperature perturbation can also be estimated from the temperature profile without island flattening. Assuming the temperature gradients to be limited by ITG in this case, we can estimate the ITG threshold to be exceeded when the island temperature reaches the temperature of the profile without island. For the H-mode profiles in Fig.~\ref{fig:temp_H_Lmode}, this would allow temperature perturbations up to $u_0 \sim 0.25$, which is consistent with the previous estimate $u_0 \sim W_N = 0.2$.

For higher temperature perturbations, turbulent transport typically leads to stiff gradients, i.e. the power required to increase the temperature beyond a certain point becomes impossibly high. Turbulent transport can thereby have significant effects on current condensation\cite{rodriguez_rf_2020}. Furthermore, turbulence was found to enhance the transport of fast electrons accelerated by EC waves in regimes where it would not greatly impact that of bulk electrons \cite{casson_effect_2015}. Thus, one aim of this section is to obtain bifurcations and hysteresis behaviour at low temperature perturbations, below the ITG threshold.

The simulations shown below were obtained by a coarse scan in the parameter space of launcher position (from an upper launcher case to halfway between upper and equatorial launcher) and launching angles (poloidal launching angle $112^\circ\leq\alpha\leq154^\circ$, measured from positive $\hat Z$, and toroidal launching angle $34^\circ\leq\beta\leq 90^\circ$, measured from negative $\hat R$ through launcher, in steps of size $1^\circ$). Those simulations not displaying a bifurcation were discarded. To trace the hysteresis curve, the power was gradually increased up to $20$~MW, the maximal EC power available in ITER, and decreased back to low powers. Close to the bifurcation, the island temperature is very sensitive to the power deposited, whence a small step size of $\sim 20$~W was used. The relative error in the island temperature at the bifurcation can be estimated as being of the same order as the relative change in $u$ in the last step before the bifurcation, which is maximally $0.4$\% in the simulations shown here.

Furthermore, the cases where the relativistic boundary (Eq.~\ref{eq:rel_constraint}) is located within the island were excluded. In such cases, the region where power can be deposited in the island is shrunk, such that higher temperature perturbations are necessary to observe a bifurcation for a given value of $\overline w_\text{eff}$. In previous work, the cases of deposition starting at the island centre and at the island edge were considered \cite{rodriguez_rf_2019}.

The island's phase is locked such that the ray goes through its O-point. Adjustment of a locked island's phase to deposit power at the island's O-point has been achieved in DIII-D with external magnetic perturbations \cite{volpe_avoiding_2015, choi_feedforward_2018, choi_simultaneous_2019}.

The EC wave propagation and damping were obtained from GENRAY\cite{smirnov_calculations_1995}, using a cold dispersion relation for the wave propagation and \citet{mazzucato_damping_1998}'s approximation of the dielectric tensor for a relativistic electron plasma for the wave damping.

\subsection{Bifurcation threshold in ITER}

\begin{figure}
\includegraphics[width=0.5\textwidth]{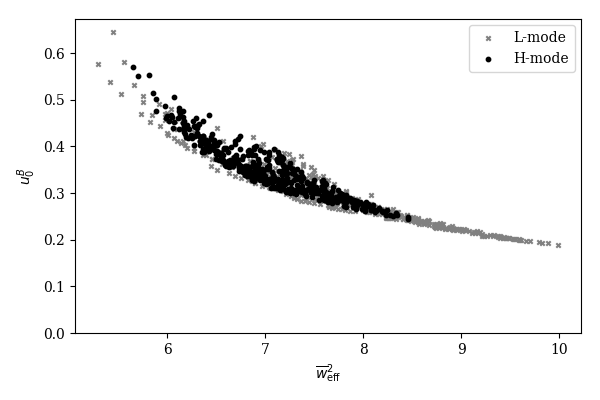}
\caption{Observed values of $\overline{w}_\text{eff}^2$ and of the normalised temperature perturbation in the island centre at the bifurcation $u_0^B$ for ITER-like H-mode and L-mode scenarios. Increasing $\overline{w}_\text{eff}^2$ helps to obtain bifurcations at lower temperature perturbations.}
\label{fig:u0_weff2}
\end{figure}

The observed values of $\overline{w}_\text{eff}^2$ (as defined in Eq.~\ref{eq:w_eff_avg}) and of the normalised temperature perturbation in the island centre at the bifurcation $u_0^B$, are shown in Fig.~\ref{fig:u0_weff2}. In the limit of very small $\overline{w}_\text{eff}^2$, the damping is already strong in the linear regime, such that nonlinear effects cannot help to focus or draw in the power deposition, hence no bifurcation is observed. When going to very large $\overline{w}_\text{eff}^2$, too little power is deposited in the island for nonlinear effects to be relevant. Thus, an intermediate region where bifurcations can be observed is found, e.g. $5.5 \lesssim \overline{w}_\text{eff}^2 \lesssim 8.5$ in the H-mode scenario (Fig.~\ref{fig:u0_weff2}). Bifurcations at higher $\overline{w}_\text{eff}^2$ can be obtained in the L-mode scenario (up to $\overline{w}_\text{eff}^2 \approx 10$), as the temperature at the island separatrix $T_s$ is lower (Fig.~\ref{fig:temp_H_Lmode}), yielding higher effective powers in the RHS of the diffusion equation (Eq.~\ref{eq:diffusion}). Analogously to \citet{rodriguez_rf_2019}, we define a normalised power density as $P_0 \equiv PW^2 \overline{w}_\text{eff}^2 / (4 V_\text{island} n T_s \chi_\perp)$, with the input wave power $P$, average density $n$, separatrix temperature $T_s$ and island volume $V_\text{island}=8\pi W r_r R_0/M$. Higher effective powers can thus be achieved e.g. by reducing the island temperature and density.

\begin{figure}
\includegraphics[width=0.5\textwidth]{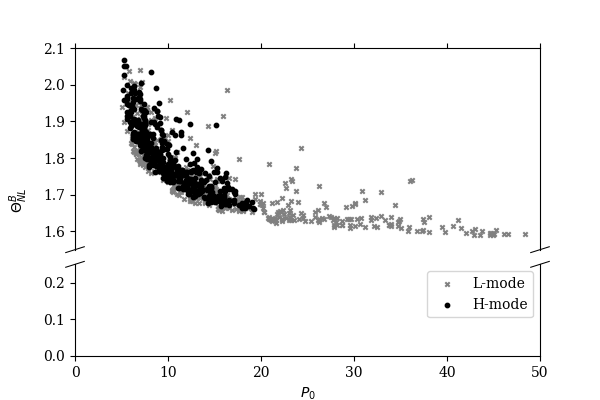}
\caption{Nonlinear amplification parameter at the bifurcation $\Theta_{NL}^B = \overline{w}_\text{eff}^2 u_0^B / (1+u_0^B)$ as a function of the normalised wave input power density $P_0$ at the bifurcation. The values of $\Theta_{NL}^B \approx 1.6-2.0$ are approximately constant, with a trend of slightly decreasing values for increasing $P_0$.}
\label{fig:theta_P0}
\end{figure}

As can be readily seen in Fig.~\ref{fig:u0_weff2}, there is a strong correlation between the temperature perturbation at the bifurcation $u_0^B$ and the $\overline{w}_\text{eff}^2$ value. The nonlinear amplification parameter at the bifurcation 
\begin{equation}
    \Theta_{NL}^B \equiv \overline{w}_\text{eff}^2 \; u_0^B / (1+u_0^B) \label{eq:theta_NL_bifurcation}
\end{equation}
 is shown in Fig.~\ref{fig:theta_P0} as a function of normalised power $P_0$. Most of the data points are in the region $\Theta_{NL}^B \approx 1.6-2.0$, with a trend of a small decrease with increasing input power $P_0$ at the bifurcation. Due to the near constancy of $\Theta_{NL}^B$, it can be effectively used as a threshold parameter, below which no bifurcation can be obtained. It is especially useful as $\overline{w}_\text{eff}^2$ is obtained from a simple ray tracing calculation in the unperturbed temperature profile. Therefore, a lower bound on the temperature perturbation necessary to observe a potential bifurcation can be obtained from solving Eq.~\ref{eq:theta_NL_bifurcation} for $u_0^B$, without solving the full nonlinear problem.

A small amount of spread in the value of $\Theta_{NL}^B$ at a given $P_0$ can be observed. Some deviation is not surprising, as the behaviour for strong variations of $w_\text{eff}$ within the island may not be fully captured by the averaged $\overline{w}_\text{eff}$ (Eq.~\ref{eq:w_eff_avg}), and the temperature dependence of $w_\text{eff}^2 \propto 1/T$ assumed in the definition of $\Theta_{NL}$ (Eq.~\ref{eq:theta_NL_def}) is only approximate.

The values of $\Theta_{NL}^B$ agree well with previous results. Using a constant power deposition and no wave depletion, \citet{reiman_suppression_2018} find that a bifurcation occurs when $\Theta_{NL}^B = w_0^2 u_0^B \approx 1.2$ for a slab model, and $\Theta_{NL}^B = w_0^2 u_0^B \approx 1.4$ for a realistic island geometry. Incorporating depletion of the wave, using a slab model and assuming an exponentially decreasing power deposition, \citet{rodriguez_rf_2019} find a bifurcation when $\Theta_{NL}^B = w_0^2 u_0^B \approx 1.2-1.5$, for deposition starting at the island edge. In comparison, the values of $\Theta_{NL}^B \approx 1.6-2.0$ from Fig.~\ref{fig:theta_P0} tend to be higher. This can be explained by our use of a realistic island geometry instead of a slab model, as a similar increase was observed in \citet{reiman_suppression_2018}. Moreover, the tendency of $\Theta_{NL}^B$ to slowly decrease with increasing input power $P_0$ observed in Fig.~\ref{fig:theta_P0} is also consistent with previous analysis \cite{rodriguez_rf_2019}.

Even though no detailed optimisation was performed in the simulations shown in this study, it can be seen from Fig.~\ref{fig:u0_weff2} that bifurcations were obtained at temperature perturbations down to $u_0 = 0.24$~and~$0.19$ for the H-mode and L-mode temperature profiles, respectively. This suggests that bifurcations could be obtained before the ITG threshold is exceeded at $u_0 \sim 0.2 - 0.25$ (Eq.~\ref{eq:ITG_threshold}) for realistic parameter values. The ITG threshold criterion was only roughly estimated, however; and more realistic calculations which include stiff-gradient effects self-consistently will thus need to be undertaken in the future. 

The lowest $u_0$ values at bifurcation were obtained for upper launcher cases, close to ITER's planned upper launcher position. In these cases, $w_\text{eff}$ tends to be approximately flat inside the island as the ray propagation in the poloidal plane occurs mostly in the $\hat{Z}$-direction, and the resonance is thus approached slowly. To put the necessary temperature perturbations into perspective, values of $u_0 \geq 0.2$ for island widths $W/a  \sim 0.2$ have been observed in TEXTOR \cite{the_textor_team_effect_2007, westerhof_tearing_2007}.

Large toroidal launching angles ($\beta \geq 34^\circ$) were chosen in this study to obtain higher $w_\text{eff}$, yielding stronger nonlinear effects, although ITER's upper launcher is planned to operate at a smaller toroidal launching angle of $20^\circ$. Moreover, single rays were used in this study, instead of gaussian beam profiles. A detailed investigation of current condensation effects for the planned ITER upper launcher steering mirrors, also incorporating gaussian beams represented by multiple rays, as well as stiff gradient effects, is left for future work.

\section{Dynamic variation of RF wave launching angles}
\label{sec:hysteresis_launching_angle}

Current condensation can lead to hysteresis as elevated island temperatures can draw in and maintain the power deposition close to the island centre, instead of having it be deposited closer to the island edge or even outside of the island. A hysteresis curve can be traced by e.g. varying the wave power or island width \cite{reiman_suppression_2018, rodriguez_rf_2019}. Another way to obtain hysteresis is to dynamically vary the toroidal or poloidal launching angle, the second of which we will demonstrate in this section.

Furthermore, we will show that the dynamic variation of the poloidal launching angle can help to circumvent the shadowing effect, a nonlinear inhibition effect \cite{rodriguez_rf_2019, jin_pulsed_2020}. At high island temperatures, a significant fraction of the wave power may be damped at the island edge before the wave reaches the island centre. This not only leads to reduced island temperatures but also to possibly destabilising currents driven close to the island separatrix. One way to bring the power deposition toward the island O-point despite the shadowing effect involves pulsing the input wave power\cite{jin_pulsed_2020}. We will show in this section that the shadowing effect can also be avoided by varying the poloidal launching angle, such that damping at the island edge is reduced. Note that similar results can be achieved by varying the toroidal launching angle.

\begin{figure}
\includegraphics[width=0.5\textwidth]{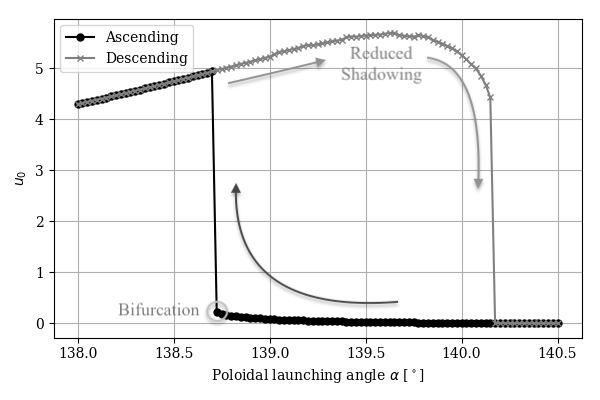}
\caption{Hysteresis in normalised temperature at island centre $u_0$ from variation of the poloidal launching angle $\alpha$. Circle data points show the ascending part of the hysteresis curve, going from the lower to the upper branch (decreasing $\alpha$), while crosses show the descending part, going from the upper branch to the lower branch (increasing $\alpha$). }
\label{fig:hyst_betast_u0}
\end{figure}

We again investigate the ITER-like H-mode case of Sec.~\ref{sec:ITER_bifurcation}, with a launcher situated close to ITER's planned upper launcher position \cite{figini_assessment_2015}, $R=7$~m, $Z=4.3$~m. Again, a $2/1$ island of width $W_N = 0.2$ (in units of the normalised radius $\rho_N$), a constant diffusion coefficient $\chi_\perp = 0.1$~m$^2$s$^{-1}$ and $20$~MW of EC wave power are considered. The toroidal launching angle is held fixed at $\beta = 50^\circ$, while the poloidal launching angle is varied from $\alpha = 140.5^\circ$ down to $138^\circ$, and back up, in small steps of $0.025^\circ$ to accurately trace the hysteresis curve. The resulting hysteresis curve for the island temperature is shown in Fig.~\ref{fig:hyst_betast_u0}.

\begin{figure}
\includegraphics[width=0.5\textwidth]{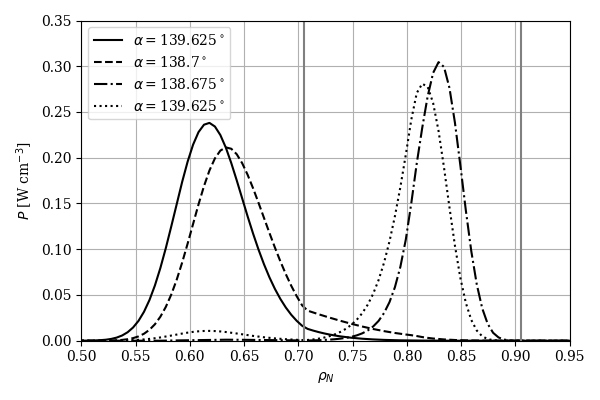}
\caption{Evolution of power deposition profile during hysteresis from variation of the poloidal launching angle. Vertical gray lines indicate the magnetic island edges. At first, almost all power is deposited outside of the island (solid curve, $\alpha = 139.625^\circ$). The power deposition is then moved into the island by decreasing the poloidal launching angle, until enough power is deposited for a bifurcation to occur at $\alpha \approx 138.675^\circ$ (dashed to dashdotted curves). Increasing the poloidal launching angle back to $\alpha = 139.625^\circ$ (dotted curve), hysteresis behaviour is displayed, as most of the power is still deposited within the island.}
\label{fig:hyst_betast_pdep}
\end{figure}

The power deposition at several points along the hysteresis curve is shown in Fig.~\ref{fig:hyst_betast_pdep}. Initially, at high poloidal launching angles, little to no power is deposited inside the island (e.g. $\alpha = 139.625^\circ$ in Fig.~\ref{fig:hyst_betast_pdep}). The poloidal launching angle is decreased, moving the power deposition into the island, until a bifurcation is reached at $\alpha \approx 138.675^\circ$. Increasing the poloidal launching angle back to $\alpha = 139.625^\circ$, the power is still deposited in the island, i.e. the system displays hysteresis behaviour.

The power deposition after the transition to the upper branch at $\alpha = 138.675^\circ$ is not centred on the island's O-point ($\rho_N \approx 0.8$), due to the aforementioned shadowing effect. Reducing the poloidal launching angle beyond the bifurcation further increases the shadowing effect, as can be seen from the decrease in the temperature on the upper branch of Fig.~\ref{fig:hyst_betast_u0}. However, by increasing the poloidal launching angle on the upper branch, the damping can be reduced at the island edge, moving the peak of the power deposition profile towards the island's O-point, as shown in Fig.~\ref{fig:hyst_betast_pdep} (dotted curve, $\alpha = 139.625^\circ$). This results in a higher central island temperature, as can be seen in Fig.~\ref{fig:hyst_betast_u0}.

The hysteresis in the poloidal launching angle shown above could be preferable to that in the EC wave power for an experimental investigation of current condensation. Compared to the hysteresis in power, it allows operation at the maximum EC power, which was shown in Fig.~\ref{fig:theta_P0} to lead to bifurcations at lower values of $\Theta_{NL}^B \equiv \overline{w}_\text{eff}^2 u_0^B / (1+u_0^B)$, i.e. bifurcations could be obtained at lower temperatures. In the case shown here, the temperature at the bifurcation is small for the H-mode case, $u_0^B = 0.26$, although no significant optimisation or large parameter scans were performed. 

Note that the upper branch solution of Fig.~\ref{fig:hyst_betast_u0} will undoubtedly be modified by stiff gradient effects, due to the large temperature perturbations ($u_0 > 4$) that will trigger ITG instabilities. However, the mechanism to circumvent shadowing presented here is more generally valid. Current condensation including stiff gradient effects has been investigated analytically\cite{rodriguez_rf_2020}, with a corresponding numerical treatment left for future work.

\section{Conclusion}

In this study, we presented a numerical treatment of current condensation effects, coupling a ray-tracing code for the wave propagation and damping, with a heat diffusion equation solver to obtain the temperature in the magnetic island. This allows to investigate current condensation in realistic scenarios, in particular including the island geometry and relativistic effects in the damping. These were shown to lead to a generalisation of the bifurcation threshold identified in previous analytical work \cite{reiman_suppression_2018, rodriguez_rf_2019}. Furthermore, bifurcations were obtained for realistic parameter values ($P = 20$~MW, $\chi_\perp = 0.1$~m$^2$s$^{-1}$) in ITER-like H- and L-mode scenarios, at low temperature perturbations $u_0 \sim 0.2$. Stiff gradient effects could be negligible at such low temperature perturbations; more detailed calculations which self-consistently include stiff gradient effects are however needed to demonstrate this. Finally, we showed that dynamically varying the poloidal launching angle can lead to a current condensation induced hysteresis and can remedy to the nonlinear inhibition brought about by the shadowing effect. Current condensation could enable the stabilisation of large magnetic islands, leading to improved disruption avoidance. Therefore, an optimisation study for ITER, including stiff gradient effects, the planned ITER launcher position and the use of multiple rays to represent gaussian beams, is in preparation.

\begin{acknowledgments}
The simulations shown in this study were run on the PPPL research cluster. This work was supported by U.S. DOE DE-AC02-09CH11466 and DE-SC0016072. The data that support the findings of this study are available from the corresponding author upon reasonable request.

\end{acknowledgments}

\appendix
\section{Derivation of diffusion equation in island geometry}
\label{sec:derivation_diff_eq}
We assume the magnetic island to be symmetric in the radial coordinate $r$, an approximation for narrow islands in a large aspect ratio tokamak of circular cross-section. We define the island flux coordinate $\sigma$, ranging from $0$ in the centre to $1$ at the separatrix, as
\begin{equation}
    r - r_r = \pm \frac{W}{2} \sqrt{ \sigma^2 - \sin^2\left ( M\zeta/2\right)}, \label{eq:island_def}
\end{equation}
with the normalised radius at the resonant surface $r_r$, the helical angle $\zeta = \theta - N\varphi / M \in [-2\pi/M, 2\pi/M)$, the poloidal (toroidal) angles $\theta$ ($\varphi$) and poloidal (toroidal) mode numbers $M$ ($N$). A new angular coordinate $\eta \in [-\pi/, \pi)$ is defined as
\begin{equation}
    \sin( M\zeta/2) = \sigma \sin(\eta). \label{eq:island_def_eta}
\end{equation}
Using the definition of $\eta$, Eq.~\ref{eq:island_def} can be rewritten as
\begin{equation}
    r - r_r = \frac{W}{2} \sigma \cos \eta.
\end{equation}
We now investigate the steady-state diffusion equation in the island,
\begin{equation}
    \nabla \cdot \left( n \chi \nabla T \right) = - p\label{eq:general_ss_diff_eq},
\end{equation}
with $\chi$ the heat diffusion coefficient, $n$ the plasma density, and $p$ the power density. Integrating over the island volume up to the flux surface $\sigma$,
\begin{align}
    \int_{-\pi}^{\pi}d\eta & \int_{0}^{2\pi}d\varphi\int_{0}^{\sigma}d\sigma'\;J\;\nabla \cdot  \left(n\chi \nabla T \right) \\
    &= - \int_{-\pi}^{\pi}d\eta \int_{0}^{2\pi}d\varphi\int_{0}^{\sigma}d\sigma'\;J\; p, \nonumber
\end{align}
where, in a large aspect ratio approximation, the Jacobian $J$ is given by
\begin{equation}
    J^{-1} = \nabla \sigma \cdot \nabla \eta \times \nabla \varphi \approx \frac{M}{W r_r R_0} \frac{\sqrt{1-\sigma^2 \sin^2{\eta}}}{\sigma}. \label{eq:jacobian}
\end{equation}
Then, assuming parallel diffusion to be significantly stronger than perpendicular diffusion, $\chi_\perp \ll \chi_\parallel$, the temperature can be assumed to be equilibrated on flux surfaces and thus becomes a function of $\sigma$ only, $T=T(\sigma)$. Defining $P_{\text{dep}}(\sigma)$ as the triple integral on the right hand side, which represents the total power deposited inside the flux surface $\sigma$, and taking the density $n$ and cross-field thermal diffusivity $\chi_\perp$ to be constant, we obtain
\begin{equation}
    \frac{\partial T}{\partial \sigma} \cdot 2\pi \int_{-\pi}^{\pi} d\eta \; J \; | \nabla \sigma| ^2 = - \frac{P_\text{dep}(\sigma)}{n\chi_\perp}. \label{eq:intermediate_diff_eq}
\end{equation}
Evaluating the integral on the left hand side for large aspect ratios, we obtain
\begin{equation}
    \int_{-\pi}^{\pi} d\eta \; J | \nabla \sigma| ^2 \approx \frac{4 R_0 r_r W}{M} \; \frac{1}{(W/2)^2} \; \frac{ E(\sigma) - \left(1-\sigma^2\right) K(\sigma)}{\sigma}. \label{eq:flux_avg_grad_rho_sq}
\end{equation}
Combining Eqs.~\ref{eq:intermediate_diff_eq}~and~\ref{eq:flux_avg_grad_rho_sq} results in the diffusion equation in the island geometry,
\begin{equation}
    \frac{\partial u}{\partial \sigma} = - \frac{P_\text{dep}(\sigma)}{n\chi_\perp T_s} 
    \frac{\sigma}{E(\sigma) - \left(1-\sigma^2\right) K(\sigma)} \frac{(W/2)^2 }{V_\text{island}}, \label{eq:diffusion_appendix}
\end{equation}
where $V_\text{island} = 8\pi R_0 r_r W / M$ is the island volume. The expression in Eq.~\ref{eq:diffusion_appendix} is equivalent to Eq.~\ref{eq:diffusion}, presented in the main text. The term containing elliptic integrals in Eq.~\ref{eq:diffusion_appendix} is an island geometric term. Previous studies of current condensation \cite{reiman_suppression_2018, rodriguez_rf_2019, rodriguez_rf_2020} used a slab model of the island, for which the diffusion equation reduces to
\begin{equation}
    \frac{\partial u}{\partial \sigma} = - \frac{P_\text{dep}(\sigma)}{n\chi_\perp T_s} 
    \frac{(W/2)^2 }{V_\text{island}}. \label{eq:diffusion_equation_slab}
\end{equation}
The equations Eq.~\ref{eq:diffusion_appendix}~and~\ref{eq:diffusion_equation_slab} are very similar, but for the added island geometric term in the former.

\section{Validity of $w_{\text{\textnormal{eff}}}$ formula}
\label{sec:validity_weff}

We compare the formula for $w_{\text{eff}}$ in Eq.~\ref{eq:w_eff2_formula} with a numerical finite differences evaluation of Eq.~\ref{eq:def_weff} using the damping coefficient from the ray-tracing code GENRAY. The damping formula by \citet{mazzucato_damping_1998} is used in the ray-tracing calculations.

\begin{figure}
\includegraphics[width=0.5\textwidth]{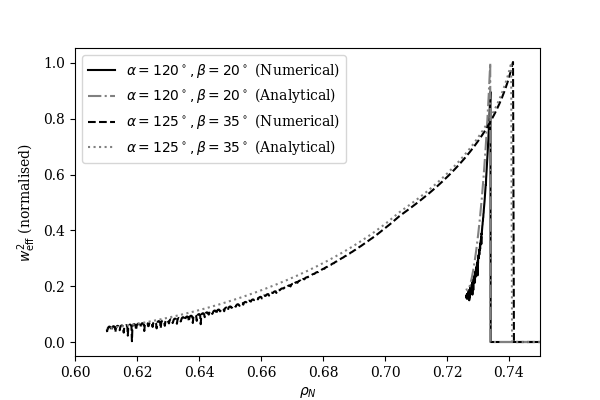}
\caption{Comparison of $w_\text{eff}$ obtained from Eq.~\ref{eq:w_eff2_formula} (Analytical) and from numerical evaluation of GENRAY damping coefficient (Numerical). Two cases are considered, at low toroidal launching angles ($\beta = 20^\circ$), with $N_\parallel \sim 0.3$, and higher toroidal launching angles ($\beta = 35^\circ$), with $N_\parallel \sim 0.5$. The $w_\text{eff}^2$ values are normalised by factors $15.62$ and $2.94$ for $\beta = 20^\circ$ and $\beta = 35^\circ$, respectively.}
\label{fig:comp_weff2_num_ana}
\end{figure}

The ITER-like H-mode profile of Sec.~\ref{sec:ITER_bifurcation} is used, and two cases with differing launching angles are considered, to obtain the case of low $N_\parallel \sim 0.3$ and that of medium $N_\parallel \sim 0.5$. The temperature in the damping regions is in the range $T\approx4-6$~keV. The resulting curves are shown in Fig.~\ref{fig:comp_weff2_num_ana}, where the $w_\text{eff}^2$ values have been normalised by factors $15.62$ and $2.94$ for $\beta = 20^\circ$ and $\beta = 35^\circ$, respectively. The agreement is excellent, with Eq.~\ref{eq:w_eff2_formula} only slightly overestimating the value of $w_\text{eff}$. Note that the relativistic constraint (Eq.~\ref{eq:rel_constraint}) is very prominent for the low $N_\parallel \sim 0.3$ case, where the wave strongly damps in a narrow spatial region, at low values of $w_\text{eff}$.

\section{Temperature dependence and classical limit of $w_\text{\textnormal{eff}}$}
\label{sec:temp_dependence_weff}

The temperature dependence of $w_{\text{eff}}$ in Eq.~\ref{eq:w_eff2_formula} is complicated due to the modified Bessel functions in the $F(\xi, a)$ term (Eq.~\ref{eq:def_F_xi}). In the remainder of this Appendix, we will show that $w_\text{eff}^2 \propto 1/T$, as was assumed e.g. in Eq.~\ref{eq:theta_NL}, is a valid approximation in the limit of $\xi \gg 1$ and when $| 1-(\Omega/\omega)^2| \sim N_\parallel^2$. If $| 1-(\Omega/\omega)^2| \ll N_\parallel^2$, the $1/T$ proportionality can still hold provided that $\mu | 1-(\Omega/\omega)| \gg 1$.

The modified Bessel functions can be expanded in the limit $\xi \gg 1$, which is sensible as $\mu \gg 1$, even for thermonuclear temperatures. However, reasonably large $N_\parallel $ and $R$ also have to be assumed for $\xi \gg 1$ to hold, the latter of which also translates to a requirement of large $N_\parallel$. Indeed, low $N_\parallel$ values will generally display significant damping close to the relativistic boundary (Eq.~\ref{eq:rel_constraint}), where $R\rightarrow0$. 

In the limit $\xi \gg 1$, the modified Bessel functions can be approximated as \cite{olver_nist_2020}
\begin{align}
    I_\nu(\xi \gg 1) &\approx \frac{e^\xi}{\sqrt{2\pi\xi}} \sum_{k=0}^{\infty}\left(-1\right)^k \frac{b_k(\nu)}{\xi^k}, \\
    b_k(\nu) &= \frac{(4\nu^2-1)(4\nu^2-3)...(4\nu^2-(2k-1)^2)}{k!8^k}. \nonumber
\end{align}
Then, Eq.~\ref{eq:def_F_xi} reduces to
\begin{align}
    &F(\xi\gg 1, a) \approx \label{eq:F_xi_approx}\\ 
    &\xi \frac{(a-1)^2 -\frac{(a-1)(3a-11)}{2\xi} + \frac{3}{2} \frac{a^2 - 10a + 13 }{\xi^2} + \frac{15a - 42}{\xi^3} }{(a-1)^2 - \frac{1}{\xi}(a-5)(a-1) + \frac{6}{\xi^2}(2-a) }. \nonumber
\end{align}
Consider the parameter $\epsilon = (a-1)$. The limit $\epsilon \rightarrow 0 $ is equivalent to approaching the resonance $Y \rightarrow 1$, where $Y \equiv \Omega / \omega$. From Eq.~\ref{eq:def_a},
\begin{equation}
    \epsilon = \frac{Y}{\sqrt{1 - \frac{Y^2 - 1}{N_\parallel^2}}} - 1, \label{eq:epsilon_formula}
\end{equation}
such that $\epsilon \ll 1$ only for $| Y^2-1| \ll N_\parallel^2$. This condition is part of the classical limit. Note that although the relativistic boundary (Eq.~\ref{eq:rel_constraint}) constrains the damping to occur for $Y^2 \geq 1-N_\parallel^2$ on the tokamak low field side, the condition $| Y^2-1| \ll N_\parallel^2$ is a much stronger constraint.

First, consider the case where $| Y^2-1| \sim N_\parallel^2$. Then, $F(\xi, a) \approx \xi$ and, from Eq.~\ref{eq:w_eff2_formula},
\begin{equation}
    -w_\text{eff}^2 \approx 1 + \mu \left(1 - \frac{Y}{1-N_\parallel^2} + \frac{N_\parallel R}{1-N_\parallel^2} \right). \label{eq:weff2_away_from_resonance}
\end{equation}
In the regime where $| Y^2-1| \sim N_\parallel^2$ and $\mu \gg 1$, the first term is negligible, such that $w_\text{eff}^2  \propto \mu \propto 1/T$, as desired.

We now treat separately the regime $| Y^2-1| \ll N_\parallel^2$. First, rewrite Eq.~\ref{eq:w_eff2_formula} as a function of $\xi$ and $a$,
\begin{equation}
    -w_\text{eff}^2 = 1 + \frac{a\xi}{N_\parallel^2} \left(\sqrt{\left(1-N_\parallel^2\right)\left(1-N_\parallel^2/a^2\right)} - 1 \right) + F(\xi, a). \label{eq:weff2_xi_a}
\end{equation}
Furthermore, Eq.~\ref{eq:F_xi_approx} can be rewritten as
\begin{equation}
    F(\xi, a=1+\epsilon) \approx \xi - \frac{1}{2} \frac{\left(\xi\epsilon \right)^2 + 12 \left(\xi\epsilon\right) + 30}{\left(\xi\epsilon \right)^2 + 4\left(\xi\epsilon\right) + 6}. \label{eq:F_xi_approx_eps}
\end{equation}
Inserting this expression into Eq.~\ref{eq:weff2_xi_a}, we obtain
\begin{equation}    
-w_\text{eff}^2 \approx \frac{\xi\epsilon^2}{2\left( N_\parallel^2 -1\right)}  + \frac{1}{2} \frac{\left( \xi\epsilon\right)^2 - 4\left(\xi\epsilon\right) - 18}{\left(\xi\epsilon\right)^2 + 4\left( \xi\epsilon \right) + 6}. \label{eq:weff_approx_close_resonance}
\end{equation}
Using $| 1-Y^2| \ll N_\parallel^2 $ and defining $\delta = 1-Y$, Eq.~\ref{eq:epsilon_formula} yields $\xi \epsilon \approx \mu \delta$ and $\epsilon \approx \delta (1-N_\parallel^2)/N_\parallel^2$. Then, Eq.~\ref{eq:weff_approx_close_resonance} reduces to
\begin{equation}    
w_\text{eff}^2 \approx \frac{\mu\delta^2}{2 N_\parallel^2}  - \frac{1}{2} \frac{\left( \mu\delta\right)^2 - 4\left(\mu\delta\right) - 18}{\left(\mu\delta\right)^2 + 4\left( \mu\delta \right) + 6}. \label{eq:weff_approx_mu_delta_close_resonance}
\end{equation}
The classical limit can then be obtained in the ordering $\mu \delta \gg 1$, such that
\begin{equation}
    w_\text{eff}^2 \approx \frac{\mu\delta^2}{2 N_\parallel^2} - \frac{1}{2} = w_0^2. \label{eq:weff_approx_classical}
\end{equation}
This indeed corresponds to the exact classical limit, obtained by inserting $\alpha \propto \mu^{-1/2} \exp{(\mu \delta^2/(2N_\parallel^2))}$ into Eq.~\ref{eq:def_weff}. Note that we previously omitted the weak $\mu^{-1/2}$ dependence of the classical damping, assuming that the exponential term would dominate, an invalid assumption when the resonance is approached, $\delta \rightarrow 0$. In this limit, $(\omega - \Omega)/(k_\parallel v_T) \rightarrow 0$, and the strong sensitivity of damping to temperature is lost as damping occurs on the bulk thermal electrons. From Eq.~\ref{eq:weff_approx_classical}, the $w_\text{eff}^2 \propto 1/T$ proportionality holds strictly in the classical limit when the first term dominates, i.e. $\mu \delta^2 / N_\parallel^2 \gg 1$. 


Note that the condition $\mu \delta \gg 1$ was required to obtain the classical limit, which was not mentioned in the original study by \citet{fidone_role_1982}. There, only the $\epsilon_{11}''$ component of the dielectric tensor was considered, which reduces to the classical damping with only the limits $\mu \gg 1$ and $N_\parallel^2 \gg | 1-Y^2| $. We are however interested in the $\epsilon_{33}''$ component, for which $\mu | 1-Y| \gg 1$ is also required, as can also be obtained by considering the square bracket term in Eq.~\ref{eq:eps33_O1_mode} directly.

\begin{figure}
\includegraphics[width=0.5\textwidth]{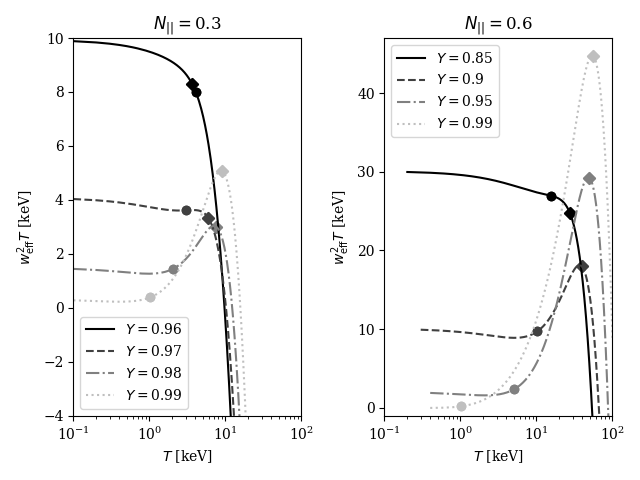}
\caption{Change of $w_\text{eff}^2 T$ as a function of temperature, for two values $N_\parallel = 0.3, 0.6$ and several values of $Y = \Omega / \omega$. Circles indicate temperatures for which $\mu (1-Y)=5$, while diamonds correspond to temperatures for which $\xi = 5$.}
\label{fig:test_weff_T}
\end{figure}

It is instructive to consider the quantity $w_\text{eff}^2 T$, as shown in Fig.~\ref{fig:test_weff_T} as a function of temperature, for different values of $N_\parallel$ and $Y = \Omega / \omega = 1-\delta$. Constancy of $w_\text{eff}^2 T$ indicates that the $w_\text{eff}^2 \propto 1/T$ proportionality holds.

First, consider the diamonds in Fig.~\ref{fig:test_weff_T}, representing the temperature values for which $\xi = 5$. It can be seen that for larger temperatures, the $\xi \gg 1$ approximation and thus also $w_\text{eff}^2 \propto 1/T$ will break down, with $w_\text{eff}^2$ even going to negative values. Away from the resonance, more specifically when $| 1-Y^2| \sim N_\parallel^2$, we should expect $\xi \gg 1$ to be the only relevant criterion, according to Eq.~\ref{eq:weff2_away_from_resonance}. Indeed, looking at e.g. $N_\parallel = 0.3$ and $Y = 0.96$, for which $| 1-Y^2| /N_\parallel^2 = 0.87$, or $N_\parallel = 0.6$ and $Y = 0.85$, for which $| 1-Y^2| /N_\parallel^2 = 0.77$, $w_\text{eff}^2 T$ is roughly constant to the left of the diamonds, where $\xi \gg 1$ is satisfied. Note that the condition $\xi \gg 1$ is more restrictive for lower $N_\parallel$ values.

Secondly, consider the circles in Fig.~\ref{fig:test_weff_T}, representing the point on each curve for which $\mu \delta = 5$. For those cases where $\xi \gg 1$ and $| 1-Y^2| \ll N_\parallel^2$, a further requirement $\mu\delta \gg 1$ is needed to reduce $w_\text{eff}$ to the classical limit. If $\mu\delta \gg 1$ is not fulfilled, the $w_\text{eff}$ of Eq.~\ref{eq:weff_approx_mu_delta_close_resonance} will have a more complicated temperature dependence. Indeed, for small $\delta = 1-Y$ (e.g. $N_\parallel = 0.6$ and $Y=0.99$, for which $| 1 - Y^2 | / N_\parallel^2 = 0.06$), the circles in Fig.~\ref{fig:test_weff_T} prove to be good indicators of a change from a regime where $w_\text{eff}^2 T$ is constant, to one where the behaviour is more complicated.

Summarising, the assumption $w_\text{eff}^2 \propto 1/T$ thus strictly holds for $\xi \gg 1$, and when simultaneously $| 1-Y^2| \sim N_\parallel^2$. If $| 1-Y^2| \ll N_\parallel^2$ however, $w_\text{eff}^2 \propto 1/T$ can still hold for $\mu | 1-Y| \gg 1$ (and more strictly also $\mu (1-Y)^2 \gg N_\parallel^2$). In the simulations shown in Sec.~\ref{sec:ITER_bifurcation}, typical values are $\mu \sim 10^2$, $N_\parallel \sim 0.5$, $\delta \sim 0.1$. Then $\xi \sim 80$ and $| 1-Y^2| / N_\parallel^2 \sim 0.75$, and we can assume $w_\text{eff}^2 \propto 1/T$ to be valid, according to Eq.~\ref{eq:weff2_away_from_resonance}.

\section{Formulas for $w_{\text{\textnormal{eff}}}$ in the X2 mode}
\label{sec:weff_X2_mode}

Whereas the O1 mode considered in Sec.~\ref{sec:relativistic_damping} is most relevant for ITER, the X2-mode is more relevant for several existing tokamaks, like DIII-D or AUG. Therefore, we repeat here the procedure in Sec.~\ref{sec:relativistic_damping} and derive a formula for $w_\text{eff}$ for the X2-mode.

Just like for the O1 mode, the components of the dielectric tensor for a relativistic electron plasma can be obtained from \citet{fidone_role_1982}. The first diagonal component is reproduced here:
\begin{equation}
    \epsilon_{11}^{\prime \prime} = \frac{\pi \omega_p^2}{2 \Omega^2} \left( \frac{R_2}{N_\parallel}\right)^{5/2} \frac{N_\perp^2 S_2}{\sqrt{1-N_\parallel^2}} I_{5/2}(\xi_2) e^{\mu \left(1-2\frac{\Omega/\omega}{1-N_\parallel^2} \right)} \label{eq:eps11_X2_mode},
\end{equation}
with
\begin{align}
    & R_2 = \sqrt{\left( \frac{2\Omega}{\omega}\right)^2 - 1 + N_\parallel^2}, \\
    & \xi_2 = \frac{N_\parallel R_2 \mu}{1-N_\parallel^2},\\
    & S_2 = H\left(  \left(\frac{2\Omega}{\omega}\right)^2 - 1 + N_\parallel^2, \right)
\end{align}
and the other symbols previously defined for Eq.~\ref{eq:eps33_O1_mode}. The damping coefficient can be approximated as 
\begin{equation}
    \alpha \approx \frac{\omega}{cN} \epsilon_{11}^{\prime \prime}.
\end{equation}
Then, using Eq.~\ref{eq:def_weff}, $w_\text{eff}^2 \approx T\partial_T (\ln{(\epsilon_{11}^{\prime \prime})})$ in conjunction with Eq.~\ref{eq:eps11_X2_mode}, we obtain
\begin{equation}
    -w_\text{eff}^2 = \mu \left( 1-2\frac{\Omega/\omega}{1-N_\parallel^2}\right) + \xi_2 \frac{I_{3/2}\left( \xi_2\right)}{I_{5/2}\left( \xi_2\right)} - \frac{5}{2}.
\end{equation}
 
\bibliography{references}

\end{document}